# Electron Paramagnetic Resonance of Mn in Bi$_2$Se$_3$ Topological Insulator

Agnieszka Wolos,[1,2] Aneta Drabinska,[2] Maria Kaminska,[2] Andrzej Hruban,[3]
Stanislawa G. Strzelecka,[3] Andrzej Materna,[3] Miroslaw Piersa,[3]
Magdalena Romaniec,[3] Ryszard Diduszko[3]

[1]Institute of Physics, Polish Academy of Sciences, Al. Lotnikow 32/46, 02-668 Warsaw, Poland.
[2]Faculty of Physics, University of Warsaw, ul. Hoza 69, 00-681 Warsaw, Poland.
[3]Institute of Electronic Materials Technology, ul. Wolczynska 133, 01-919 Warsaw, Poland.

Electron paramagnetic resonance was used to investigate Mn impurity in Bi$_2$Se$_3$ topological insulator grown by the vertical Bridgman method. Mn in high-spin S = 5/2, Mn$^{2+}$ configuration, was detected regardless of the conductivity type of the host material. This means that Mn$^{2+}$(d$^5$) energy level is located within the valence band, and Mn$^{1+}$(d$^6$) energy level is outside the energy gap of Bi$_2$Se$_3$. The electron paramagnetic resonance spectrum of Mn$^{2+}$ in Bi$_2$Se$_3$ is characterized by the isotropic g-factor |g| = 1.91 and large axial parameter D = -4.20 GHz × h. This corresponds to the zero-field splitting of the Kramers doublets equal to 8.4 GHz × h and 16.8 GHz × h, respectively, which is comparable to the Zeeman splitting for the X-band. Mn in Bi$_2$Se$_3$ acts as an acceptor, effectively reducing native-high electron concentration, compensating selenium vacancies, and resulting in p-type conductivity.

1. Introduction.

Doping topological insulators[1,2] with transition metal ions is expected to lead to novel physical phenomena; magnetic impurities are likely to modify topological surface states, in particular causing band gap opening at the Dirac point.[3,4] It has been also indicated that Dirac electronic states mediate the RKKY interaction among localized magnetic moments and can result in long-range ferromagnetic order.[3] Mn is one of the most popular magnetic dopants because it has both the highest solubility of all transition metals and usually high-spin electronic configuration. Recently, ferromagnetism has been found in Mn-doped molecular beam epitaxy-grown Bi$_2$Se$_3$; its origin either from bulk or surface states has been considered.[5,6,7] The transport response of surface states in the presence of magnetic ions has been also reported.[8] In view of often contrasting results assigning ferromagnetism to either bulk or surface states, indicating paramagnetic, spin-glass or ferromagnetic phases,[5,6,7,9,10,11,12,13] it becomes clear that the investigation into magnetism in topological insulators is still in the initial stage and requires an in-depth study. Although surface magnetism gained the greatest interest, the magnetic properties of the bulk need to be clarified for the complete understanding of transition metal-doped topological insulators, the surface states of which are coupled with bulk states, thus forming one energy structure.

The way in which a Mn impurity is incorporated in the crystal lattice of topological insulators has not been fully explored yet. The lattice site location, charge and spin states of the impurity are particularly important for the clarification of the origins of magnetic interactions. The spin state of Mn in Bi$_2$Se$_3$ and in particular its dependence on the Fermi level has not been investigated experimentally yet. In general, it has been believed that Mn in Bi$_2$Se$_3$ is in the 2+ valence state and the impurity acts as an acceptor. The acceptor character has been determined based on analysis of the influence of Mn doping on the electric transport properties of Bi$_2$Se$_3$.[14]

In this work we report on the first Electron Paramagnetic Resonance (EPR) spectra of Mn in Bi$_2$Se$_3$, pointing to Mn$^{2+}$(d$^5$) configuration in both n- and p-type samples. The EPR technique is a powerful tool for studying transition metal elements in classical semiconductors as well as in topological insulators. Here, the said technique has provided a fingerprint of the Mn$^{2+}$ configuration. It has been confirmed that the Mn ion substitutes for the Bi site in Bi$_2$Se$_3$, occurs in high-spin S = 5/2 configuration regardless of the conductivity type and takes on the axial symmetry of the Bi$_2$Se$_3$ crystal lattice. Influence of Mn doping on the electric conductivity and the crystal morphology of Bi$_2$Se$_3$ is also discussed.

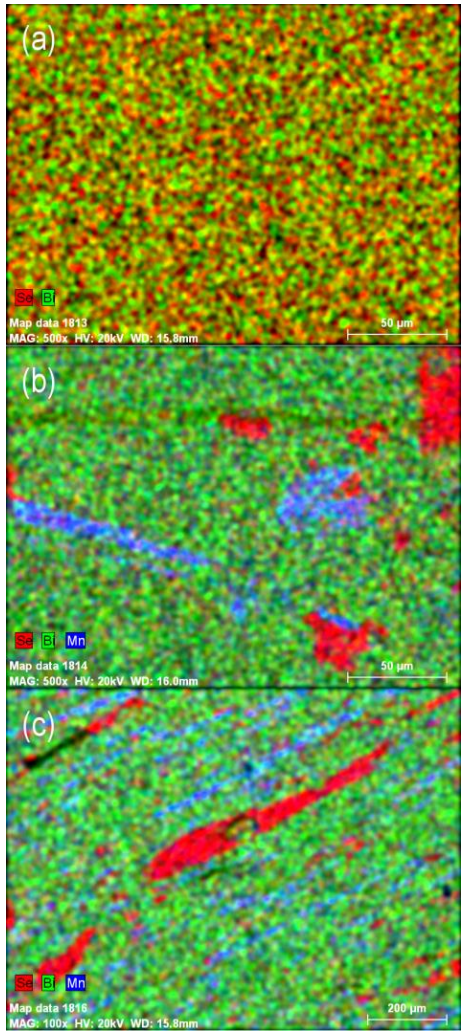

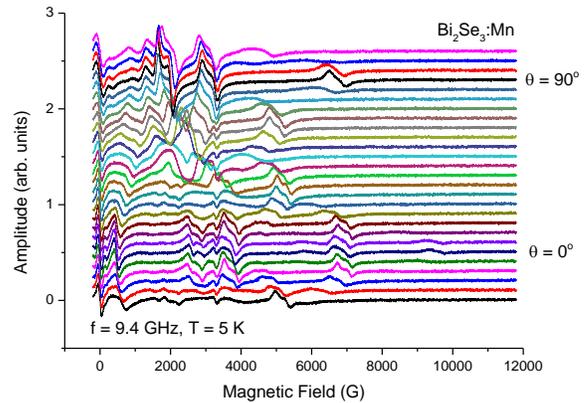

Figure 2. The anisotropy of the $Mn^{2+}$ EPR signal in $Bi_2Se_3$. The magnetic field was rotated in the plane between the $Bi_2Se_3$ trigonal c-axis (θ = 0) and the bisectrix s-axis (θ = 90). Sample was p-type, ingot #2.

Figure 1. EDX maps of $Bi_2Se_3$ doped with Mn, ingot #5, (a) 28 mm from the seed, homogeneous distribution of the elements, (b) 92 mm and (c) 126 mm from the seed - precipitates of Se and Mn are visible (metallic Se and $MnSe_2$ from the X-rays). The precipitates intercalate van der Waals gaps.

2. Samples and the Experiment.

Samples for the EPR studies were grown by the vertical Bridgman method. The growth details can be found in Ref. 15. The crystal growth process consisted of the following three stages: synthesis, directional crystallization and annealing. The process was carried out in quartz ampoules sealed under vacuum of $10^{-5}$ Tr. High purity (99.999 % – 99.9999 %) elements (Bi, Se, and Mn) were charged into ampoules. At the first stage of synthesis a slow melting of the components occurred and transitional phases, $Bi_2Se$ and $BiSe$, were created at temperatures below 600°C. After increasing the temperature to about 710°C, $Bi_2Se_3$ was formed. The batch was homogenized by mechanically moving the ampoules, annealed at the temperature of about 820°C, and gradually cooled down to room temperature. Crystallization from the synthesized batch was performed in a vertical Bridgman furnace. At first, a slow melting of the batch occurred at the temperature above 710°C. When the temperature reached 820-840 °C, the ampoule was lowered towards the furnace having the axial temperature gradient of 12-18 °C/cm, at a constant rate of 1-3 mm/h. After crystallization, the ampoule was annealed at the temperature of 500-550 °C. The complete process lasted about 90 hrs.

The obtained ingots, usually measuring 100 mm in length and up to 20 mm in diameter, contained large grains of a regular quintuple structure, with an area of up to 100 $mm^2$ and 100 mm long. The trigonal c-axis was perpendicular to the growth direction, while the binary n-axis stayed parallel to it. Due to segregation effects, the defect and dopant concentration and therefore the carrier concentration varied along the ingot. Hence, to compare the ingots, representative samples were cut at about 30 mm from the seed for all the grown ingots, unless otherwise indicated.

Electrical characterization was performed by the Hall method in the van der Pauw configuration. Structural characterization was performed using X-ray diffraction and Scanning Electron Microscopy

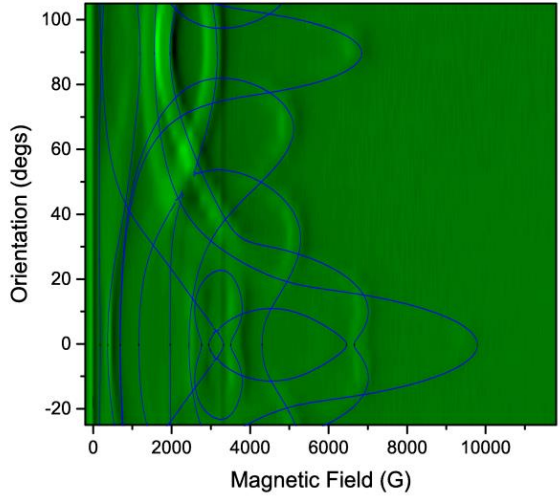

Figure 3. The anisotropy of the X - band EPR signal of $Mn^{2+}$ in $Bi_2Se_3$ (the intensity map) compared to the position of resonances simulated using the XSophe software with the effective spin Hamiltonian parameters $|g| = 1.91$ and $D = -4.20$ GHz (blue solid lines). 0 degs means the magnetic field parallel to the $Bi_2Se_3$ c-axis, while 90 degs stays for the magnetic field parallel to the bisectrix s-axis. p-type sample from the ingot #2.

(Auriga Zeiss, equipped with an Energy Dispersive X-ray EDX analyser).

Electron Paramagnetic Resonance studies were performed using a Bruker Elexsys E580 spectrometer equipped with the X-band (f = 9.5 GHz) $TE_{102}$ resonance cavity and Oxford continuous-flow cryostat. The temperature was lowered down to 5 K, whereas the magnetic field reached up to 17 000 G. Due to the use of magnetic field modulation and the lock-in technique, the recorded spectra represent the first derivative of microwave absorption.

### 3. Electrical and Structural Properties of $Bi_2Se_3$:Mn.

$Bi_2Se_3$ crystallizes in a rhombohedral lattice with a unit cell composed of three quintuple layers, each containing five alternating bismuth (Bi) and selenium (Se) hexagonal planes, stacked along the c-axis in the following atomic order: Se-Bi-Se-Bi-Se. Quintuple layers are bonded by weak van der Waals forces (dominant) with the so-called van der Waals gaps created between them, thus determining natural cleavage planes. The outermost planes in the quintuple layer composed of selenium atoms tend to host a large number of selenium vacancy defects

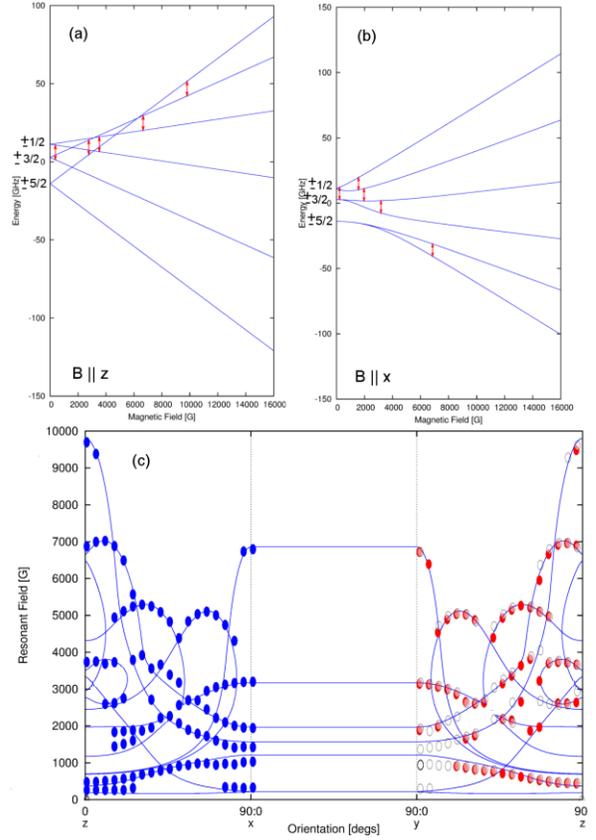

Figure 4. (a) and (b) the splitting of the $Bi_2Se_3$:$Mn^{2+}$ ground state in the magnetic field calculated using the XSophe simulation software with the effective spin Hamiltonian parameters $|g| = 1.91$ and $D = -4.20$ GHz. The hyperfine structure is excluded for clarity. Red arrows indicate the most intense transitions in the X - band EPR. The zero-field splitting equals to $|2D| = 8.4$ GHz × h and $|4D| = 16.8$ GHz × h. (c) The anisotropy of the $Bi_2Se_3$:$Mn^{2+}$ EPR signal. Open points - n-type sample from the ingot #2, closed points - p-type sample from the same ingot. Solid blue lines are simulated using the XSophe software with the effective spin Hamiltonian parameters listed above. x corresponds to the direction of the $Bi_2Se_3$ binary n-axis, y – the bisectrix s-axis, z - the trigonal c-axis.

acting as double donors, $V_{Se}$. Therefore, undoped $Bi_2Se_3$ is typically n-type, with actual electron concentration strongly dependent on growth conditions. In the Bridgman growth method, it can be effectively controlled by varying the stoichiometry of the melt. Crystals grown from the stoichiometric melt show room-temperature electron concentration as high as $10^{19}$-$10^{20}$ $cm^{-3}$, while increasing the selenium-to-bismuth ratio lowers electron concentration to ~ $10^{17}$ $cm^{-3}$. It has been reported that Mn in $Bi_2Se_3$ acts

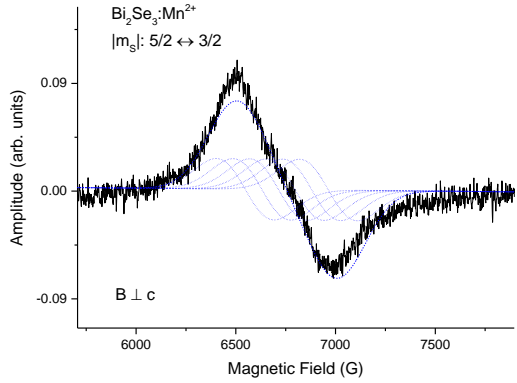

Figure 5. Decomposition of the $Mn^{2+}$ resonance line originating from the transition between $|m_S| = 5/2$ and $|m_S| = 3/2$ levels on the six hyperfine components (dotted blue lines). The magnetic field is perpendicular to the $Bi_2Se_3$ c-axis. The distance between the hyperfine structure lines is 83 G.

as an acceptor, and eventually converts conductivity to p-type.[14]

First experiments focused on doping $Bi_2Se_3$ with Mn were performed using a stoichiometric melt (bismuth-to-selenium-to-manganese atomic ratio Bi:Se:Mn = 1.85:3.00:0.15). The obtained ingots were n-type with electron concentration (at room temperature) as high as $1.1 \times 10^{20}$ cm$^{-3}$, which is higher than in the case of undoped material grown under the same conditions (i.e. $2 \times 10^{19}$ cm$^{-3}$). A similar effect was reported in Ref. 11 and attributed to the influence of Mn-doping on the concentration of native defects. It was postulated that Mn decreases the concentration of the $Bi_{Se}$ antisite defect acting as an acceptor. However, it has been recently argued that the $Bi_{Se}$ defect is actually a donor defect that is located in the outermost Se layer (near the van der Waals gap) and can be viewed as a $V_{Se}$ and Bi interstitial ($V_{Se}$ is a double donor and Bi remains neutral).[16] Thus the simple explanation of $Bi_{Se}$ concentration reduced by Mn doping cannot be used anymore, and the reason for the increase of electron concentration is not clear. X-ray diffraction and EDX analyses have revealed the presence of a BiSe transitional phase and metallic Mn precipitates, which may suggest problems with crystallization. We have encountered a similar situation when doping $Bi_2Se_3$ with a Ca acceptor. Growth from the stoichiometric melt has not ended in obtaining p-type material because the CaSe phase has crystallized before the crystallization of $Bi_2Se_3$, causing shortage of Se and an almost total loss of Ca from the melt. Therefore, our further crystallization experiments were carried out using the non-stoichiometric melt, containing a slight or moderate Se excess. This approach has proved to be a compromise between lower electron concentration of the base (undoped) material and good crystal morphology (avoiding Se precipitation), and in addition it has ensured successful acceptor (Ca) doping.[15]

Table 1 shows the growth conditions of Mn-doped $Bi_2Se_3$ versus its carrier concentration. Free electron concentration in the case of undoped material for a specific stoichiometry of the melt is also given for comparison. The ingots grown from the melt containing a slight Se excess (Bi:Se:Mn = 1.80:3.15:0.050) showed a variable type of conductivity along the ingot; n-type and p-type regions occurred alternately along the ingot. The n-type regions had the electron concentration of the order of $9 \times 10^{18}$ cm$^{-3}$, while the p-type regions had the hole concentration of the order of $(6-9) \times 10^{18}$ cm$^{-3}$. It will be shown below that both the n-type and p-type regions contained similar $Mn^{2+}$ concentration, of the order of $1.5 \times 10^{19}$ cm$^{-3}$.

A moderate Se excess in the melt (Bi:Se = ~1.6:3.30) led to uniform p-type conductivity along the ingot, with hole concentration systematically increasing with increasing the nominal content of Mn in the melt from $3.6 \times 10^{18}$ cm$^{-3}$ for 0.015 atomic parts up to $1.1 \times 10^{19}$ cm$^{-3}$ for 0.100 atomic parts of Mn. The experiments including material doping with both Mn and Ca acceptors were also carried out. The composition of the melt Bi:Se:Mn:Ca = 1.70:3.30:0.008:0.008 allowed lowering electron concentration to $4.6 \times 10^{18}$ cm$^{-3}$, while the composition of 1.69:3.30:0.0037:0.0035 lowered it even more, i.e. to $6 \times 10^{17}$ cm$^{-3}$ at room temperature.

It was discussed in our earlier publication, Ref. 15, that growth from the melt containing a Se excess (moderate or high) as well as doping are associated with the formation of precipitates of foreign phases. They usually intercalate in the van der Waals gaps of $Bi_2Se_3$. Figure 1 presents EDX images of samples grown from the melt with a moderate Se excess (#5) cut at different distances from the seed. The initial (seed) sections of the ingot were free of precipitates, while the distal (tail) sections showed precipitation of Se and Mn. The X-ray diffraction analysis identified foreign phases as metallic Se and $MnSe_2$, which is shown in Table 1. The occurrence of Mn and Se precipitates in distal sections of the ingot suggests the enrichment of the melt with these elements during growth. Enrichment or loss of one or more of the melt components during growth is typical for the Bridgman technique, results from the growth thermodynamics, and is often described by a so-called segregation coefficient.

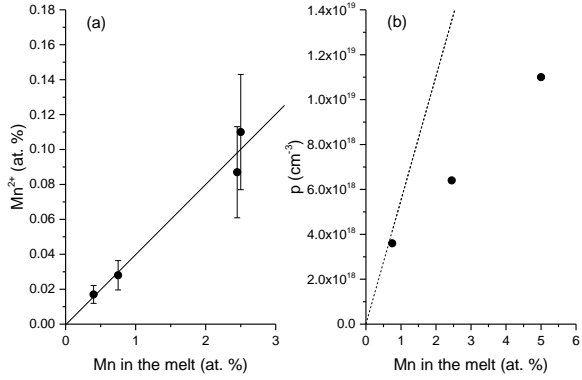

Figure 6. (a) $Mn^{2+}$ concentration determined from the EPR as a function of the nominal content of Mn in the melt. Solid line is a guide to the eye. About 4% of the Mn incorporated as substitutional isolated impurity in the $Bi_2Se_3$ lattice. (b) hole concentration as a function of the nominal content of Mn in the melt for a series of samples grown with moderate Se excess. Dashed line is for $Mn^{2+} = p$ ($Mn^{2+}$ calculated as 4% of the nominal Mn concentration).

### 4. Electron Paramagnetic Resonance of Mn in $Bi_2Se_3$.

Figure 2 shows the X-band EPR spectra collected for a p-type sample obtained from the ingot #2. The magnetic field was rotated in the plane between the trigonal c-axis ($\theta = 0$) and the bisectrix s-axis ($\theta = 90$) of $Bi_2Se_3$. Figure 3 shows the same set of spectra in the form of an intensity map, together with the simulated anisotropy of the resonances. The simulation will be described later in the text. The signal visible at the zero-magnetic field in Fig. 2 is a magnetoresistance feature originating from the weak antilocalization phenomenon, which is similar to that observed in epitaxial graphene.[17,18,19] It will not be further discussed in this paper. The rest of the lines visible in Fig. 2 originate from the Mn impurity in $Bi_2Se_3$. The spectrum is not typical for $Mn^{2+}$ configuration in the axial crystal field. It usually consists of five fine structure lines, each split into a hyperfine sextet. When the magnetic field is parallel to the c-axis the distance between the hyperfine lines is the highest, while at about 50 deg. from that direction the lines merge.[20,21] Here, the resonance does not resemble the well-known pattern due to large zero-field splitting of the $Mn^{2+}$ ground state, comparable to the energy of the applied microwaves.

A Mn atom has $[Ar]3d^54s^2$ electronic configuration. When substituting the Bi site in $Bi_2Se_3$ ($Bi^{3+}$ ion), three electrons are moved to the bonds. The Mn is left in $[Ar]3d^4$ configuration. However, $^6A_1$ configuration without the orbital momentum is the most preferential one and entails the acceptance of an extra electron on the Mn d-shell from the $Bi_2Se_3$ conduction band. The Mn impurity is then left in $[Ar]3d^5$ configuration ($Mn^{2+}$). The half-filled d-shell carries the high-spin $S = 5/2$. In $Bi_2Se_3$ the Mn impurity has the $C_{3v}$ point symmetry and can be effectively described by the following spin Hamiltonian (taking into account the axial symmetry):[22]

$$\hat{H} = \mu_B \vec{B}\hat{g}\vec{S} + D\left[S_z^2 - \frac{1}{3}S(S+1)\right] + \vec{S}\hat{A}\vec{I},$$

Eq.1

where $\hat{g}$ is the tensor describing the effective electron g-factor, $\mu_B$ is the Bohr magneton, D is the axial crystal-field parameter determining the zero-field splitting of the $Mn^{2+}$ ground state. $\vec{S}$ is the electron spin with $S_z$ being its projection on the trigonal c-axis ($S = 5/2$), $\vec{I}$ is the nuclear spin ($I = 5/2$ of the natural isotope $^{55}Mn$). $\vec{S}\hat{A}\vec{I}$ term describes the hyperfine structure parameterized by the hyperfine constant A. Higher order a and F parameters are reduced from the Hamiltonian because $|D| \gg |a-F|$.

The ground state of the $Mn^{2+}$ ion is split by the axial crystal-field into three Kramers doublets separated by the 4D and 2D energy, respectively. The magnetic field splits the Kramers doublets; the ground state of $Mn^{2+}$ is composed of six energy levels, each characterized by the $m_s$ quantum number, $m_s = -5/2, -3/2, \ldots, +3/2, +5/2$ (in a high magnetic field). Each level is split into six components, due to the hyperfine interaction with the Mn nucleus having nuclear spin $I = 5/2$. The transitions allowed in the EPR experiment are characterized by $\Delta m_s = 1$ and $\Delta m_I = 0$. In a high magnetic field the selection rules lead to the resonance spectrum consisting of five lines of the fine structure, having the intensity ratio in the first approximation equal to 5:8:9:8:5, each of them split into six hyperfine components. In the case of $Mn^{2+}$ in $Bi_2Se_3$, parameter D is comparable to the energy of the microwaves (the Zeeman splitting for the X-band). The resonance transitions occur in the low magnetic field regime, where $m_s$ is no longer a good quantum number due to the mixing of the states, and the resonance pattern is no longer regular. The calculated structure of the $Mn^{2+}$ ground state in the magnetic field is illustrated in Figs 4a and 4b with corresponding resonance transitions marked with arrows. The scheme of the energy levels was obtained using the XSophe software (from Bruker)[23] with parameters provided later in the text.

Table 1. Growth details (nominal melt composition), room temperature carrier concentration, free electron concentration in a reference crystal with no Mn doping, $Mn^{2+}$ concentration determined from EPR, and the morphology (determined from X-ray diffraction) of $Bi_2Se_3$:Mn samples grown by the vertical Bridgman method. Due to segregation effects both the carrier concentration and the Mn concentration vary along the ingot. The given numbers refer to the regions at about 1/3 of the ingot length; samples were cut about 30 mm from the seed, except for in the case of sample #8, which was cut close to the tail of the ingot.

| Ingot number | Nominal melt composition (atomic ratio) | | Comments | Carrier concentration ($cm^{-3}$) | | Reference n ($cm^{-3}$) | $Mn^{2+}$ | Morphology |
|---|---|---|---|---|---|---|---|---|
| | Bi:Se | Mn | | n | p | | | |
| #1 | 1.85:3.00 | 0.150 7.5 at. % | Stoichiometric melt | $1.1 \times 10^{20}$ | | $2 \times 10^{19}$ | No EPR signal | Tail: BiSe, Mn |
| #2 | 1.80:3.15 | 0.050 2.5 at. % | Slight Se excess n/p regions | $8.8 \times 10^{18}$ | $5.8 \times 10^{18}$ | $5 \times 10^{18}$ | $1.5 \times 10^{19}$ $cm^{-3}$ 0.11 at. % | No data |
| #3 | | | | $8.6 \times 10^{18}$ | $8.8 \times 10^{18}$ | | | No data |
| #4 | 1.60:3.30 | 0.100 5 at. % | Moderate Se excess | | $1.1 \times 10^{19}$ | $1 \times 10^{18}$ | No EPR signal | Seed: Se, $MnSe_2$ Tail: Se, $MnSe_2$ |
| #5 | 1.65:3.30 | 0.049 2.45 at. % | | | $6.4 \times 10^{18}$ | | $1.2 \times 10^{19}$ $cm^{-3}$ 0.087 at. % | Seed: No Tail: Se, $MnSe_2$ |
| #6 | 1.68:3.30 | 0.015 0.75 at. % | | | $3.6 \times 10^{18}$ | | $3.9 \times 10^{18}$ $cm^{-3}$ 0.028 at. % | Seed: No Tail: Se |
| #7 | 1.70:3.30 | 0.008 0.4 at. % | Moderate Se excess + Ca doping (0.008) | $4.6 \times 10^{18}$ | | No data | $2.4 \times 10^{18}$ $cm^{-3}$ 0.017 at. % | No data |
| #8 | 1.69:3.30 | 0.0037 0.185 at. % | Moderate Se excess + Ca doping (0.0035) | $6.0 \times 10^{17}$ | | No data | No EPR Signal | No data |

Figure 4c shows the anisotropy of the resonance spectra, both measured and simulated using XSophe. Only the fine structure lines were marked. The magnetic field was rotated in two planes, between the $Bi_2Se_3$ c- and the n-axis and between the c- and the s-axis. Both the simulation and the experiment indicate isotropy of the signal when the magnetic field is rotated in the plane perpendicular to the c-axis (between the n- and the s-axis). The position of the resonance lines for both the n-type and the p-type samples follows the same angular anisotropy, regardless of the conductivity type. The following parameters best reflect the anisotropy of the signal: the isotropic g-factor $|g| = 1.91$ and $|D| = 4.20$ GHz × h. Sign of parameter D determines the order of the energy levels, and thus the intensity of the respective transitions. Using the Xsophe, parameter D can be determined as negative, D = -4.20 GHz × h. The hyperfine structure cannot be resolved from the spectra, because the width of the hyperfine lines is higher than the distance between them. However, a fit of the six hyperfine components to the recorded resonance line can be still performed, yielding the hyperfine line separation of the order of 83 G, thus giving the hyperfine parameter A of the order of A = 83 G × $g\mu_B$ = 232 MHz × h (Fig. 5), which is a typical value for $^{55}Mn^{2+}$ in chalcogenides, see Table 2.

Concentration of substitutional $Mn^{2+}$ in $Bi_2Se_3$ samples was determined using the spin-counting software delivered by Bruker, adhering to the method based on double integration of the resonance spectra. The concentration determined in this way ranged from $3.9 \times 10^{18}$ cm$^{-3}$ up to $1.5 \times 10^{19}$ cm$^{-3}$, which corresponds to 0.028 at. % and 0.11 at. %, respectively. $Mn^{2+}$ concentrations for specific samples are listed in Table 1. The accuracy of determination of manganese concentration values did not exceed 30 %. The main error originated from uncertainty in the determination of the total intensity of the EPR transition, because the resonance spectra consisted of a number of weak resonance lines spread in a wide range of the magnetic field. Nevertheless, the double integration procedure provided the reference concentration of $Mn^{2+}$.

Only about 4 % of the nominal Mn content in the melt was substitutionally incorporated in the $Bi_2Se_3$ crystal lattice, as illustrated in Fig. 6(a) and Table 1. The melt was systematically enriched with Mn during growth, which led to the gradient of Mn concentration along the ingot, more prominent towards the tail. Therefore in some cases Mn precipitates formed in the distal sections, while the initial sections were free of them (Fig.1).

Figure 6(b) shows hole concentration in a series of samples grown with moderate Se excess (#4, #5, #6). The undoped material grown under the same conditions had electron concentration lowered to about n = $1 \times 10^{18}$ cm$^{-3}$, thus the concentration of $V_{Se}$ was significantly reduced with this stoichiometry of the melt. Hole concentration increased systematically when increasing nominal Mn content in the melt, and hence increasing the $Mn^{2+}$ concentration in the $Bi_2Se_3$ crystal lattice (Fig. 6(a)) which is consistent with the acceptor character of the Mn. However, the $Mn^{2+}$ concentration equals to the hole concentration only for low Mn doping (compare with the dashed line in Fig. 6(b) calculated as 4 % of the nominal Mn content in the melt). For higher Mn doping, the pronounced saturation of hole concentration takes place, thus suggesting that Mn not only acts as an acceptor but also influences the concentration of native defects and, in particular for high concentrations, increases the concentration of native donors. $Mn^{2+}$ resonance for the highest-doped sample (#6) was not observed. We believe that it is due to interactions between Mn ions due to reduced distance between them, that lead to the broadening of the resonance lines and prevent from observation of the resonance.

5. Discussion

A question of magnetic ordering in transition metal-doped topological insulators has been addressed and dealt with both theoretically and experimentally. It has been recently shown that Mn doping allows obtaining ferromagnetic phases in nanometer-thick molecular beam epitaxy-grown $Bi_2Se_3$, having both the surface and bulk components.[6,5,7] Doping at the level of ~ 1 at. % leads to bulk ferromagnetism with the critical temperature of the order of 6 K, for which conduction electron-mediated ferromagnetism is considered the most plausible origin of the ferromagnetic phase. The surface ferromagnetism has the critical temperature reaching up to 100 K.

The carrier mediated model of ferromagnetism, so far successfully used in a number of diluted magnetic semiconductors, requires the coexistence of transition metal ions in high-spin configuration with concentration reaching up to 2-5 at. % together with high concentration of electric charge carriers, preferably holes, mediating in the interaction.[24] Our studies have shown that Mn in $Bi_2Se_3$ occurs in S = 5/2 configuration and coexists with either n- or p-type carriers. This means that $Mn^{2+}(d^5)$ energy level is located within the valence band, and $Mn^{1+}(d^6)$ energy level is outside the energy gap of $Bi_2Se_3$. High concentration of both carrier types in $Bi_2Se_3$ can be obtained by applying specific growth conditions. However, the natural p-type results from the acceptor character of Mn in $Bi_2Se_3$. Bulk ferromagnetism was not observed in this study (no ferromagnetic resonance present) due to low concentration of Mn, reaching maximum ~ 0.1 at.

Table 2. Effective spin Hamiltonian parameters for substitutional $Mn^{2+}$ in chalcogenides. If two values for the g-factor are given, they correspond to $g_\parallel$ and $g_\perp$, respectively.

| $Mn^{2+}$ | Comments | \|g\| | \|D\| (GHz × h) | \|A\| (MHz × h) | Ref. |
|---|---|---|---|---|---|
| $Bi_2Se_3$ | Layered host | 1.91 | 4.20 | 232 | This work |
| $Cd_2P_2Se_6$ | Layered host | 2.01 2.05 | 8.57 | 210 | 25 |
| $Cd_2P_2S_6$ | Layered host |  | 1.09 |  | 25 |
| $CdGa_2Se_4$ |  | 2.0029 2.0039 | 2.75 | 182 | 26 |
| $CdGa_2S_4$ |  | 2.0012 2.0016 | 0.67 | 192 | 26 |
| $MgTe_2$ |  | 2.015 | 25.7 | 189 | 27 |

%. We focused here on the clarification of the nature of substitutional Mn in $Bi_2Se_3$ and grown samples with a low Mn concentration and good crystal morphology. Samples doped with higher Mn concentration (#1 and #4) showed presence of considerable amounts of precipitates, $MnSe_2$ or metallic Mn, and deterioration of the crystal structure which manifested in appearance of small grains. Interestingly no $Mn^{2+}$ resonance was detected in these two samples. In the case of sample #1 (the highest content of Mn in the melt) there is a doubt whether Mn incorporated substitutionally as isolated center at all. It becomes clear that non-equilibrium growth is required in order to obtain high-quality precipitate-free samples with Mn concentration at the at. % level.

The most striking result of this study is the high value of the axial parameter D, which is comparable to the Zeeman splitting for the X-band (|D| = 4.20 GHz × h). Parameter D is extremely sensitive to the local environment. Both the covalency of the metal-ligand bond and the lattice dimensionality have been found to influence its magnitude,[25] which for chalcogenides leads to the value of D reaching a few GHz. High values of parameter D have been reported for substitutional $Mn^{2+}$ in $CdGa_2S_4$ and $CdGa_2Se_4$ (0.674 GHz × h and 2.75 GHz × h, respectively)[26] as well as for $Cd_2P_2S_6$ and $Cd_2P_2Se_6$ (1.09 GHz × h and 8.57 GHz × h, respectively),[25] increasing from sulfide to selenide due to a more covalent character of the Mn-Se bond. Parameter D in the case of $Mn^{2+}$ in $MgTe_2$ is even larger (25.8 GHz × h)[27], which is consistent with the further increase of the manganese-chalcogenide bond covalency (the electronegativity of the chalcogenides increases in the following order: S > Se > Te). Parameter D is expected to get higher when the lattice dimensionality is reduced. A record parameter D has been achieved in the planar environment of a porphyrin ring for $Fe^{3+}$, 600-700 GHz × h.[28] In crystalline materials the layered structure has been claimed to increase the covalency of the Mn-S bond in $Cs_2Zn_3S_4$ (|D| = 2.67 GHz × h) relative to the same bond in three-dimensional binary sulfide ZnS (|D| = 0.393 GHz × h) and $CdGa_2S_4$ (|D| = 0.674 GHz × h).[29,26] In conclusion, both the covalency of the Mn-Se bond and the layered structure of $Bi_2Se_3$ promote a high value of parameter D, as illustrated in Table 2.

We should discuss at this point the possibility that $Mn^{2+}$ creates a complex with $V_{Se}$. This in principle could happen in $Bi_2Se_3$, because $Mn^{2+}$ is a negatively charged center that can electrostatically attract positively charged selenium vacancy. Moreover, Mn substitutes Bi in the lattice thus its nearest neighbor is a selenium site. For defect complexes, a strong axial field is expected that leads to high value of the D parameter (e.g. , for the interstitial $Mn^{2+}$-indium pair in silicon the D equals to 15.6 GHz × h[30]). Such a defect complex is essentially different from the isolated impurity in the same material. It differs in the value of the D parameter and also the direction of the symmetry axis may be different. Weather the spectrum shown in Fig. 2 can originate from the defect complex? This cannot be a priori excluded, however is very unlikely. First, the $Mn^{2+}$-$V_{Se}$ complex would be a donor. $V_{Se}$ donates two electrons, while $Mn^{2+}$ accepts only one electron from the conduction band. Figs. 6(a) and (b) show that the increase of the $Mn^{2+}$ concentration (calculated from the intensity of the resonance lines) is followed by the increase of the hole concentration. If the EPR spectrum originated from the $Mn^{2+}$ coupled to $V_{Se}$ this should not take place. However, we cannot exclude completely the presence of the $Mn^{2+}$-$V_{Se}$ complex, but if it exists it is out of the resonance condition of our experiment.

The g-factor determined in this work to be g = 1.91 is only slightly reduced from the free electron value ($g_0$ = 2.0023), $\Delta g = g - g_0$ = -0.09, due to the localized nature of the Mn d-shell. For comparison, the spin resonance of conduction electrons in $Bi_2Se_3$ yields the anisotropic g-factor with $g_\parallel$ = 27.5, and $g_\perp$ = 19.5,[31] the high value of which results from the delocalization of the conduction electron wavefunction stretching over the $Bi_2Se_3$ crystal lattice containing heavy Bi element, thus experiencing strong spin-orbit coupling. The g-factor of the localized $Mn^{2+}$ center in pure ionic crystals is typically reduced by $\Delta g$ of the order of -0.0004 due to the admixture of the excited $^4$P-state into the $^6$S-ground state of the $Mn^{2+}$ ion. In chalcogenides, both positive and negative, but always relatively high $\Delta g$ has been found, i.e. delta $\Delta g$ = -0.0009 for $CdGa_2S_4$, +0.0011 for $Cd_2Ga_2Se_4$, +0.013 for $MgTe_2$, +0.03 for $Cd_2P_2Se_6$, which can be explained by the covalent mixing of the paramagnetic S-state of the Mn ion with the orbitals of the ligands, see discussion in 26 and references therein. In the covalence model $\Delta g$ scales linearly with the Pauling's covalence parameter c normalized to the number n of the ligands (c/n = 19.5%, 20.1%, and 21.8% for Mn-X bond in tetrahedrally coordinated n = 4 sites, X = S, Se, Te). Taking into account the Mn site octahedrally coordinated to six selenium atoms in $Bi_2Se_3$, c/n = 13.4 %. The negative sign of $\Delta g$ = -0.09 falls well within the trend described in Ref. 26; however, the magnitude is much higher than predicted by the covalence model, suggesting that the interaction beyond the nearest neighbors has to be included.

The doping of $Bi_2Se_3$ with Mn showed to effectively reduce native-high electron concentration, eventually leading to p-type conductivity. The hole concentration as high as ~ $1 \times 10^{19}$ cm$^{-3}$ was obtained. The lowest electron concentration was reached when both Mn and Ca-doping were combined with growth from the non-stoichiometric melt. The reduction of electron concentration down to $6 \times 10^{17}$ cm$^{-3}$ at room temperature was achieved.

6. Conclusions.

The electron paramagnetic resonance experiment proved that Mn in $Bi_2Se_3$ substitutes for a Bi site and is in high-spin $Mn^{2+}$ configuration, S = 5/2, both in n-type and p-type samples. This means that $Mn^{2+}(d^5)$ energy level is located within the valence band, and $Mn^{1+}(d^6)$ energy level is outside the energy gap of $Bi_2Se_3$. In the crystal field of $Bi_2Se_3$ the Mn ion becomes axially symmetric, and can be described by the effective spin Hamiltonian with the following parameters: |g| = 1.91 and D = -4.20 GHz × h. The zero-field splitting of the Kramers doublets is then 2D = 8.4 GHz × h and 4D = 16.8 GHz × h, which is comparable to the energy of the X-band (9.5 GHz). A high value of parameter D is typical for chalcogenides due to the covalent character of the Mn-chalcogenide bond (Mn-Se in our case).

Mn in $Bi_2Se_3$ acts as an acceptor, effectively reducing native-high electron concentration, compensating selenium vacancies, and resulting in p-type conductivity. Carrier concentration can be controlled to a significant degree by combining Mn-doping with varied stoichiometry of the melt and eventually performing co-doping with a Ca acceptor. However, one has to focus on preventing precipitates of foreign phases that can readily intercalate into van der Waals gaps.

ACKNOWLEDGEMENTS

This work was supported by the National Science Centre (Poland) funds granted under research project no. 2011/03/B/ST3/03362. We would like to thank K. Staszkiewicz for his help with the elaboration of raw data.

References:

[1] M. Z. Hasan and C. L. Kane, Rev. Mod. Phys. **82**, 3045 (2010).

[2] Haijun Zhang, Chao-Xing Liu, Xiao-Liang Qi, Xi Dai, Hong Fang, and Shou-Cheng Hang, Nat. Phys. **5**, 438 (2009).

[3] Qin Liu, Chao-Xing Liu, Cenke Xu, Ciao-Liang Qi, and Shou-Cheng Hang, Phys. Rev. Lett. **102**, 156603 (2009).

[4] Chengwang Niu, Ying Dai, Meng Guo, Yandong Ma, Baibiao Huang, and Myung-Hwan Whangbo,. J. Mat. Chem. C **1**, 114 (2013).

[5] S.-Y. Xu M. Neupane, C. Liu, D. Zhang, A. Richardella, L. A. Wray, N. Alidoust, M. Leandersson, T. Balasubramanian, J. Sanchez-Barriga, O. Rader, G. Landolet, B. Slomski, J. H. Dil, J. Osterwalder, T.-R. Chang, H.-T. Jeng, H. Lin, A. Bansil, N. Samarth, and M. Z. Hasan, Nature Physics **8**, 616 (2012).

[6] H. J. von Bardeleben, J. L. Cantin, D. M. Zhang, A. Richardella, D. W. Rench, N. Samarth, and J. A. Borchers, Phys. Rev. B **88**, 075149 (2013).

[7] D. Zhang, A. Richardella, D. W. Rench, S. Y. Xu, A. Kandala, T. C. Flanagan, H. Beidenkopf, A. L. Yeats, B. B. Buckley, P. V. Klimov, D. D. Awschalom, A. Yazdani, P. Schiffer, M. Z. Hasan, and N. Samarth, Phys. Rev. B **86**, 205127 (2012).

[8] J. G. Checkelsky, J. Ye, Y. Onose, Y. Iwasa, and Y. Tokura, Nature Physics **8**, 729 (2012).

[9] J. Choi, W. Lee, B. S. Kim, S. Choi, J. Choi, J. H. Song and S. Cho, J. Appl. Phys. **97**, 10D324 (2005).

[10] Y. S. Hor, P. Roushan, H. Beidenkopf, J. Seo, D. Qu, J. G. Chekelsky, L. A. Wray, D. Hsieh, Y. Xia, S. Y. Xu D. Qian, M. Z. Hasan, N. P. Ong, A. Yazdani, and R. J. Cava, Phys. Rev. B **81**, 195203 (2010).

[11] P. Janicek, C. Drasar, P. Lostak, J. Vejpravova, and Y. Sechovsky, Physica B **403**, 3553 (2008).

[12] Z. Salman, E. Pomjakushina, V. Pamjakushin, A. Kanigel, K. Chashka, K. Conde, E. Morrenzoni, T. Proschka, K. Sedlak, and A. Suter, arXiv:1203.4850.

[13] H. Ji, J. M. Allred, N. Ni, J. Tao, M. Neupane, A. Wray, S. Xu, M. Z. Hasan, and R. J. Cava, Phys. Rev. B **85**, 165313 (2012).

[14] Y. H. Choi, N. H. Jo, K. J. Lee, H. W. Lee, Y. H. Jo, J. Kajino, T. Takabatake, K.-T. Ko, J.-H. Park, and M. H. Jung, Appl. Phys. Lett. **101**, 152103 (2012).

[15] A. Hruban, S. G. Strzelecka, A. Materna, A. Wołoś, E. Jurkiewicz-Wegner, M. Piersa, W. Orłowski, W. Dalecki, M. Kamińska, and M. Romaniec, J. Cryst. Growth **407**, 63 (2014).

[16] F.-T. Huang, M.-W. Chu, H. H. Kung, W. L. Lee, R. Sankar, S.-C. Liou, K. K. Wu, Y. K. Kuo, and F. C. Chou, Phys. Rev. B **86**, 081104(R), (2012).

[17] A. Drabinska, A. Wolos, M. Kaminska, W. Strupiński, and J. M. Baranowski, Phys. Rev. B **86**, 045421 (2012).

[18] A. Drabinska, M. Kaminska, A. Wolos, W. Strupinski, A. Wysmolek, W. Bardyszewski, R. Bozek, and J. M. Baranowski, Phys. Rev. B **88**, 165413 (2013).

[19] A. Wolos and A. Drabinska, J. Cryst. Growth **401**, 314 (2014).

[20] T. Graf, M. Gjukic, M. Hermann, M. S. Brandt, and M. Stutzmann, Phys. Rev. B **67**, 165215 (2003).

[21] A. Wolos and M. Kaminska, *Magnetic Impurities in Wide Band-gap III-V Semiconductors,* Edited by T. Dietl, D.D. Awshalom, M. Kaminska et al., Spintronics, Semiconductors and Semimetals, Vol. 82, 325 (2008).

[22] A. Abragam and B. Bleney, *Electron Paramagnetic Resonance of Transition Ions*, Clarendon Press, Oxford 1970.

[23] G. R. Hanson, K. E. Gates, Ch. J. Noble, M. Griffin, A. Mitchell, S. Benson, J. Inorganic Biochemistry **5**, 903 (2004).

[24] T. Dietl, H. Ohno, F. Matsukura, J. Cibert, and D. Ferrand, Science **287**, 1019 (2000).

[25] D. A. Cleary, A. H. Francis, and E. Lifshitz, Chemical Physics **106**, 123 (1986).

[26] M. Schlaak and A. Weiss, Z. Naturforsch. **27a**, 1624 (1972).

[27] O. Okada and T. Miyadai, Japan. J. Phys. **17**, 231 (1978).

[28] J. E. Bennett, J. F. Gibson, and D. J. E. Ingram, Proc. Roy. Soc. **A240**, 67 (1957).

[29] M. Heming and G. Lehmann, Z. Naturforsch. **38a**, 149 (1983).

[30] J. Kreissl, W. Gehlhoff, P. Omling, and P. Emanuelsson, Phys. Rev. B **42**, 1731 (1990).

[31] A. Wolos, A. Drabinska, S. Szyszko, M. Kaminska, S. G. Strzelecka, A. Hruban, A. Materna, and M. Piersa, *The Physics of Semiconductors*, AIP Conf. Proc. 1566, 197 (2013).